# Current-induced suppression of superconductivity in tantalum thin films at zero magnetic field

Yongho Seo[1*], Yongguang Qin[1], Kyusang Choi[1,2], and Jongsoo Yoon[1]
[1]Department of Physics, University of Virginia, Charlottesville, VA 22903, U.S.A.
[2]Department of Physics, Sunchon National University, Suncheon, Jeonnam 540-742, Korea.

We report our findings on the mechanism of current-induced suppression of superconductivity in amorphous tantalum films as thin as 5nm under zero applied magnetic field. Our results indicate that the applied current generates magnetic vortices threading the films, and the dynamics of these vortices leads to the suppression of the superconductivity. Our findings also imply that the motion of current-generated magnetic vortices gives rise to nonlinear transport of which characteristics closely resemble those expected in Kosterlitz-Thouless theories with strong finite size effects.

Perhaps the most dramatic effect of an applied current to a superconductor is the full suppression of the superconductivity. For a superconductor in thin film geometry under a magnetic field above its lower critical field ($B>B_{c1}$) where the applied magnetic field penetrates the film in the form of localized tubes of magnetic flux, known as magnetic vortices, the current-induced suppression of the superconductivity is understood in the context of vortex dynamics [1-5]. An applied current exerts Lorentz force on the vortices which is balanced by viscous drag force in a steady state. In such a driven vortex system at temperatures ($T$) near the superconducting transition temperature ($T_c$), suppression of superconductivity via a dynamic instability is predicted in a model by Larkin and Ovchinnikov (LO) [1] and experimentally well established [2-4]. In the LO theory, at large vortex velocities the electric field due to vortex motion results in decreasing size of vortex cores because quasiparticles accelerated by the electric field can reach energies above the superconducting energy gap and diffuse away from the vortex core augmenting the quasiparticle population in the surrounding superconducting region. The reduction in the vortex core size causes a reduction in the viscous drag force, and the vortex motion becomes unstable and runs away to a higher velocity until the system reaches the normal conducting state. At $T<<T_c$ where electron-electron scattering is more rapid than electron-phonon scattering, it has been shown that a dynamic instability arises as the vortex expands rather than shrinks, and viscous drag is reduced because of softening of gradients of the vortex profile rather than a removal of quasiparticles [5].

However, at zero applied magnetic field (B=0), the mechanism of current-induced suppression of the superconductivity in thin films has not been clearly identified. In this Letter, we report a dynamic instability observed in superconducting amorphous tantalum films as thin as 5nm at B=0 at $T \lesssim T_c$. The instability appears as a discontinuous and hysteretic voltage jump at which the superconductivity is abruptly quenched or appears. Transport characteristics near the instability indicate that the instability arises from the dynamics of vortices generated by the bias current. Our results imply that the mechanisms for current-induced suppression of superconductivity in thin films under zero and finite-magnetic fields share a common origin.

Our samples are amorphous tantalum films, dc-sputter deposited on polished quartz substrates at a rate of 0.1 nm/sec, and patterned into a Hall bar shape (1mm × 5mm) using a shadow mask. The substrates were rotated during the deposition to ensure homogeneity. The $T_c$'s of our films in the thickness range from 2nm to 50nm are found to decrease continuously towards 0K with decreasing thickness, which is characteristic of amorphous structure of a superconducting thin film [6]. We have studied films with thicknesses, 5nm, 10nm, and 36nm. All three samples show the same main features that we discuss in this Letter, but the data presented are from the 36nm thick film. Using standard expressions for superconductors in the dirty limit [7], the zero temperature penetration depth of our 36nm thick film is calculated to be $\lambda = 1.05 \times (\rho_N/T_c)^{1/2} \approx 1.6\mu m$, the BCS coherence length $\xi(T=0) \approx 11$nm, the Ginzburg-Landau parameter $\kappa \approx 90$, the lower critical field $B_{c1}(T=0) \approx 12$mT, the upper critical field $B_{c2}(T=0) \approx 1.2$ T, normal resistivity $\rho_N \approx 2.5\ \mu\Omega$ m, and $[(dB_{c2})/(dT)]|_{T_c} \approx -2.6$ T/K. These numbers are typical for a low-$T_c$ superconducting thin film.

In figure 1**a**, the resistive superconducting transition of the 36nm thick film at B=0 is shown. Even though the $T_c$ is suppressed to ~1K from its bulk value of 4.5K [8], the transition remains very sharp demonstrating that the film is highly homogeneous. Shown in figure 1**b** are the current-biased V-I curves. As illustrated in the inset of figure 1**a**, the voltages are continuously monitored at each bias current, and the V-I traces are obtained from the equilibrium voltages. It is apparent that, with decreasing $T$ below $T_c$ a portion of a V-I curve becomes nonlinear. With further decreasing $T$, the nonlinearity grows and eventually evolves into a hysteretic voltage jump. Depending on whether the V-I traces exhibit hysteretic voltage jumps, temperatures below $T_c$ can be grouped into high and low temperature regimes. In the high temperature regime (HTR), the traces are continuous and reversible. Here the superconductivity is suppressed gradually over a range of bias current ($I_{bias}$) where the V-I's are nonlinear. In the low temperature regime (LTR), we observe hysteretic voltage jumps at which superconductivity is abruptly quenched or appears. The voltage jump is truly discontinuous indicating that it corresponds to an electronic instability because no steady state with an equilibrium voltage in the range covered by the jump is observed even with current steps of 5nA. Such discontinuous and hysteretic V-I traces of superconducting thin films at B=0 have not been observed so far.

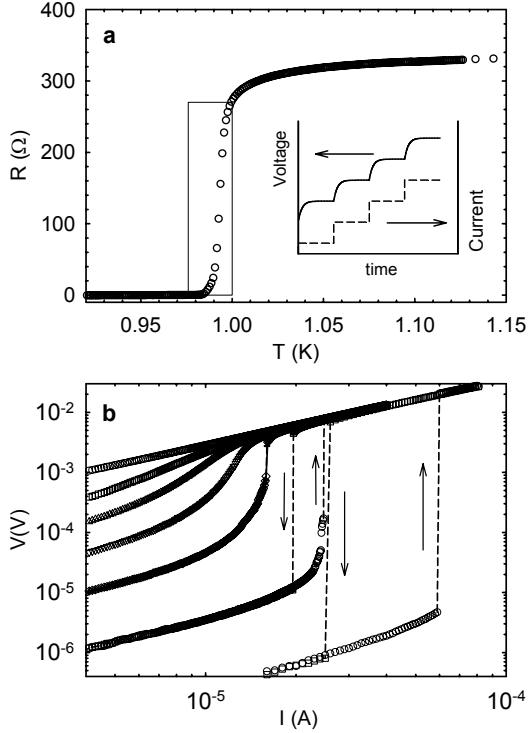

**FIG. 1. a** The resistance is measured with a 1μA dc current at B=0. A rectangle indicates the temperature range where the V-I curves (panel b) are taken. The inset illustrates our measurement methods. **b** The temperature for each trace is, from the top, 1.000K, 0.994K, 0.992K, 0.990K, 0.988K, 0.984K, and 0.976K. For the top four traces, the traces measured by increasing and decreasing currents are identical, and only current-increasing branch is plotted. For the bottom three traces, both current-increasing (open circles) and decreasing branches (open squares) are plotted. The dashed lines are to indicate discontinuous voltage jumps in the direction marked by arrows.

We first focus on the origin of the sudden quenching of the superconductivity in LTR. One may argue that the observed voltage jumps can arise from Joule-heating effect, which could elevate the sample temperature substantially above the bath temperature due to the limited thermal conduction between the sample and its thermal bath. If the sample temperature reaches the transition region where the sample resistance sharply increases with increasing $T$, the Joule heating effect can cause a thermal run-away phenomenon at a constant bias current [9, 10]. In this scenario the power at the onset of the voltage jump is expected to be independent of applied magnetic fields. Shown in figure 2**a** and 2**b** are the critical current $I_c$ and critical voltage $V_c$ at the onset of the voltage jumps on the current-increasing branch (bottom right corner of the hysteresis in figure 1**b**) as a function of magnetic field at $T$ = 0.970K. While $I_c$ decreases gently with increasing magnetic field, $V_c$ sharply rises by more than two orders of magnitude up to ~10G and then almost saturates. As shown in figure 2**c**, the critical power $P_c = I_c \times V_c$ increases about 200 times with increasing the applied magnetic field from zero to ~10G. Such a strong magnetic field dependence of the $P_c$ for B<10G clearly excludes the Joule heating scenario as a possible origin of the hysteretic voltage jumps at B=0. The hysteretic nature of the voltage jumps rules out BCS pair breaking mechanism as their possible origin.

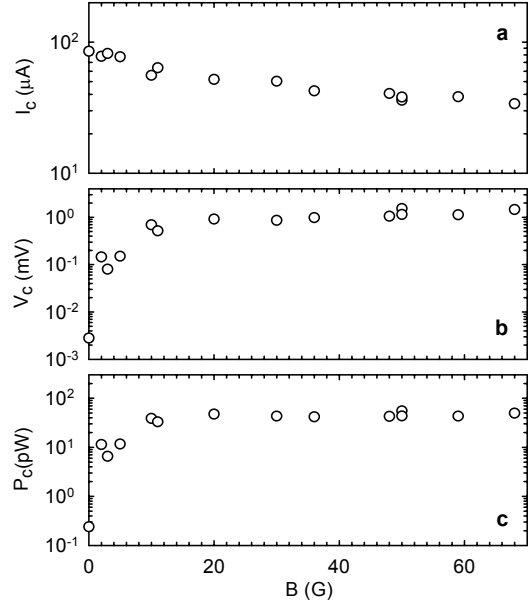

**FIG. 2.** At $T$= 0.970K, the critical current $I_c$, the critical voltage $V_c$, and the critical power $P_c$ on the increasing current branch are shown as a function of applied magnetic fields in panels a, b, and c, respectively. $I_c$ is the bias current where a hysteretic voltage jump is observed, and $V_c$ is the highest equilibrium voltage before the voltage jump occurs.

A clue to the origin comes from the close resemblance of the observed hysteretic voltage jumps to those observed in superconducting films under magnetic fields [2]. We interpret that the applied current generates magnetic vortices threading the film, and the LO-type dynamics of these current-generated vortices gives rise to the hysteretic voltage jumps. In this picture, the current-driven superconducting transition corresponds to a first-order transition, and hysteresis is a natural consequence because the onset of the superconductivity in the decreasing current branch requires nucleation of vortices. Below we describe two other observations also pointing that magnetic vortices are at work in our system at B=0.

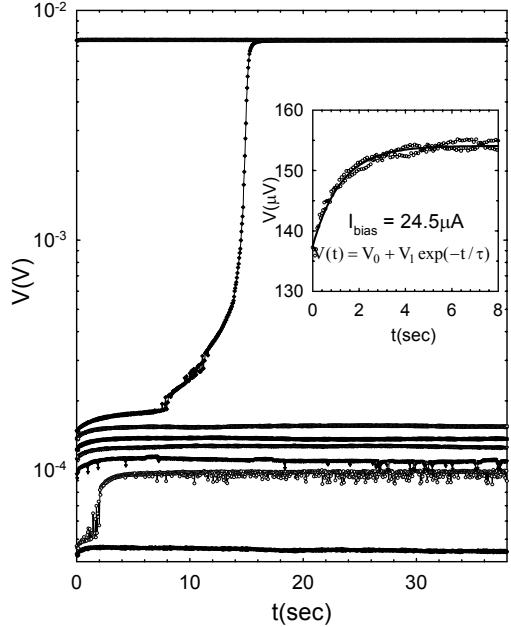

**FIG. 3**. The temperature 0.984K and the bias currents are, from 24.0 μA (bottom) to 24.7 μA with 100nA steps. The near vertical voltage rise at t ≈ 14sec in the trace of $I_{bias}$ =24.6μA marks the current-driven transition to the normal conducting state. The inset is a blow up of the initial voltage rise at $I_{bias}$= 24.5 μA, and the solid curve is a fit to the exponential function shown.

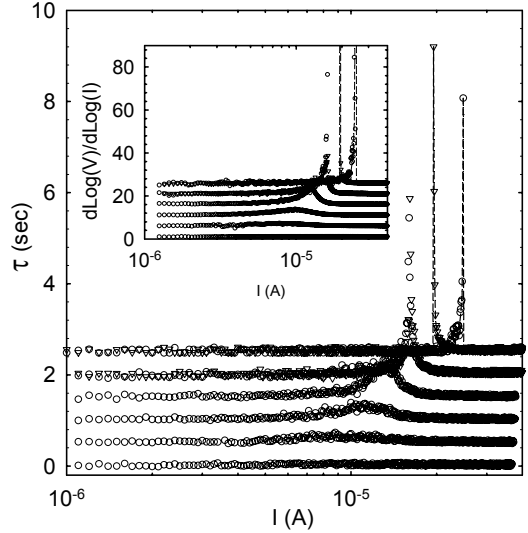

**FIG. 4.** The traces are successively shifted by 0.5 sec. The instrument response time of ~100 msec was subtracted. The inset shows numerically calculated differential V-I curves in log-log scale. Each trace is successively shifted by 5 units of the scale. For both the main panel and the inset, the temperature of each trace is, from the top, 0.984K, 0.988K, 0.990K, 0.992K, 0.994K, and 1.000K. For the 0.984K and 0.988K traces, both current increasing (open circles) and decreasing branch (open triangles) are plotted. At higher temperatures, traces of both branches coincide and only current increasing branch is shown. The dashed lines are to guide an eye.

In a driven vortex system, intermittent excess voltage noise can arise from complicated vortex configuration fluctuations due to the interplay of driving and pinning force [11, 12]. This is shown in figure 3 where the time traces of the voltage at *T*=0.984K are plotted. Intermittent excess noise is evident in the traces for $I_{bias}$ = 24.1μA, 24.2μA, and 24.6 μA. The excess noise around $I_{bias}$ = 24.1μA is observed only in a narrow range of bias current of ~150nA in our measurements with 5nA current steps (not shown). Excess noise was not measurable at $I_{bias}$ ≤ 24.0 μA, and gradually disappeared with increasing $I_{bias}$ beyond 24.2 μA. The occurrence of the excess noise only in such a narrow range of bias current clearly indicates that the noise comes from the sample. It is not until $I_{bias}$ = 24.6 μA, where the transition to the normal conducting state occurs at t ≈ 14 sec, that an excess noise becomes visible again at t ≈ 8 sec and 10 sec. In repeated measurements after thermal cyclings to *T* above 10K, sometimes with different current step sizes, the excess noise at $I_{bias}$ ≈ 24.1μA and the two separate bursts of excess noise followed by the near vertical voltage rise to the normal conducting state were always observed at the nominally same bias currents. Although the times when the near vertical voltage rise occurs were varied from ~5 sec to ~34 sec in different runs, the voltage values where the near vertical voltage rise and the accompanying two bursts of excess noise occur were the same within the experimental resolution. In the vortex dynamics picture, such a reproducibility in voltage is expected as the finite voltage response arises from the motion of vortices which dissipates energy.

Lastly, the voltage response of our system is found to be strongly dynamic. As shown in the inset of figure 3, the initial voltage rise is well described by an exponential function. The parameter τ is defined as the time constant for voltage equilibrium. Shown in figure 3 are the time constant traces extracted by fitting the time-dependent voltages for the first 8 sec to the exponential function with three fitting parameters, $V_o$, $V_1$ and τ. The results and quality of the fit are nearly the same for 1 parameter fitting where values of $V_o$ and $V_1$ are determined from the measured equilibrium voltages. At $I_{bias}$ where the transition to the normal conducting state occurs, the time constant is determined from the maximum rate of voltage rise. It should be pointed out that, within the scatter of the data, the time constants are measured to be the same by decreasing the current steps from 100nA (shown) to 5nA.

In LTR (top two traces) the time constants exhibit a diverging behaviour as the current-driven superconducting transition is approached. At $T$=0.988K, where the transition is barely hysteretic, the diverging time constants are almost symmetric as those in HTR. At lower $T$ where the transitions are strongly hysteretic, the diverging time constants are also hysteretic coinciding with the transition. Time constants as long as ~1 minute have been observed. Such a strong dynamic aspect of the voltage response in LTR can be understood in the context of vortex dynamics, where the voltage represents the average velocity of vortices. A change in bias current forces the system to establish a new steady state where the driving and drag forces are balanced. With approaching the instability, the balance becomes subtler and the time to reach a new steady state grows.

Now we turn to HTR where V-I traces are continuous and reversible. In figure 4, it is clearly visible that a peak in time constant develops with decreasing $T$. The peak grows sharper and smoothly evolves into a hysteretic and diverging peak as the $T$ is cooled to LTR. A voltage response with a time constant as long as several seconds is a manifestation that the nature of the voltage response in the peak region in HTR is also strongly dynamic as in LTR near the voltage jumps. The smooth evolution of the peak structure from the high to the low temperature regime is a clear indication that the same mechanism is behind the voltage response in the peak region in HTR and near the voltage jumps in LTR. Shown in the inset of figure 4 is the slopes of the V-I curves in log-log scale. It is evident that the differential V-I and the time constant traces exhibit similar peak structures at the same bias currents at a given $T$. Noting that a peak in a differential V-I in HTR corresponds to the nonlinear portion of the V-I where the superconductivity is gradually suppressed, our arguments lead to the conclusion that the mechanism of the current-induced suppression of the superconductivity in HTR is the same as that in LTR which we have identified as dynamics of the current-generated vortices. Unlike in LTR where the current-induced suppression of the superconductivity appears as a hysteretic voltage jump when the velocity of the vortices reach the "critical" velocity, in HTR the superconductivity gradually disappears before the motion of the vortices reaches a "critical" velocity for the LO-type instability without exhibiting a hysteretic voltage jump.

Our interpretation that the nonlinear transport in HTR arises from the dynamics of current-generated vortices has a significant implication to the traditional understanding that the nonlinear V-I's of superconducting thin films at B=0 are due to Kosterlitz-Thouless (KT) mechanism [13-16]. In the KT picture, the superconducting transition corresponds to a thermodynamic instability of "geometric" vortices, which are geometric disorder in the order parameter induced by thermal fluctuation [13]. The KT vortices are expected to exist in vortex-antivortex pairs in the superconducting phase, and a KT transition, which is a continuous phase transition, is driven by unbinding of the KT vortex pairs. In this picture, due to the current-induced KT vortex pair dissociations the transport is expected to follow a power law, V∝I$^α$, where the exponent α should exhibit a sharp jump from 1 to 3 at $T_c$ and grows bigger with further decreasing $T$ [14]. The KT picture has been challenged by experiments reporting the absence of the expected power law and the universal jump in the exponent α at $T_c$ [17, 18]. However, it has been argued that those observations could be explained within the KT framework if strong finite size effects are assumed [19, 20].

In LTR, the observed sharp hysteresis with diverging time constants for voltage response clearly rules out the possibility that the transport near the hysteretic voltage jumps are due to the KT mechanism. In HTR, on the other hand, the appearance of the nonlinear transport characteristics is similar to what is expected by the finite size KT theories, although the time constant as long as several seconds seems to be too long. However, our observations strongly indicate that the nonlinear transport in HTR arises from the dynamics of current-generated vortices, not from the KT mechanism. It should be pointed out that the KT theory, which assumes the zero current limit, may not be applicable in the current regime $\gtrsim$10 μA (or $\gtrsim$3×10$^5$ A/m$^2$) where the transport in our systems are nonlinear. Microscopic mechanisms of how a bias current generates magnetic vortices are not clear at present, but possibilities include "effective" magnetic vortices induced by non-uniform current densities caused by unavoidable inhomogeneity of the samples and generation of magnetic vortex pairs by straight current paths.

Authors acknowledge informative discussions with E. Kolomeisky and H. Fertig. This work is supported by National Science Foundation and Jeffress Memorial Foundation.

* Present address: Department of Physics and Astronomy, Northwestern University, Evanston, IL 60208, U.S.A.

[1] A. I. Larkin and Yu. N. Ovchinnikov, Sov. Phys. JETP **41**, 960 (1976).
[2] A. V. Smiolov, et al, Phys. Rev. Lett **75**, 4118 (1995).
[3] W. Klein, et. al., J. Low Temp. Phys. **61**, 413 (1985).
[4] S. G. Doettinger, et. at., Phys. Rev. Lett **73**, 1691 (1994).
[5] M. N. Kunchur, et. al., Phys. Rev. Lett. **87**, 177001 (2001).
[6] A. M. Goldman and N. Marković, Physics Today **Nov.**, 39 (1998).
[7] P. H. Kes and C. C. Tsuei, Phys. Rev. **B 28**, 5126 (1983).
[8] M. Scherschel, et. al., Physica **B 194-196**, 2289 (1994).
[9] Z. L. Xiao, et. al., Phys. Rev. **B 58**, 11185 (1998).
[10] J. Viña, et. al., Phys. Rev. **B 68**, 224506 (2004).
[11] T. Tsuboi, et.al., Phys. Rev. Lett. **80**, 4550 (1998).
[12] A. Kiliç, et. al., Physica **C 384**, 321 (2003).
[13] J. M. Kosterlitz and D. J. Thouless, J. Phys. **C 6**, 1181 (1973).
[14] B. I. Halperin and D. R. Nelson, J. Low Temp. Phys. **36**, 599 (1979).
[15] A. F. Hebard and A. T. Fiory, Phys. Rev. Lett. **50**, 1603 (1983).
[16] S. A. Wolf, et. al., Phys. Rev. Lett. **47**, 1071 (1981).
[17] M. M. Rosario, et. al., Phys. Rev. **B 61**, 7005 (2000).
[18] J. M. Repaci, et. al., Phys. Rev. **B 54**, R9674 (1996).
[19] S. T. Herbert, et. al., Phys. Rev. **B 57**, 1154 (1998).
[20] K. Medvedyeva, et. al., Phys. Rev. **B 62**, 14531 (2000).